\documentclass[a4paper,11pt]{article}
\pdfoutput=1 % if your are submitting a pdflatex (i.e. if you have
\usepackage{jheppub} % for details on the use of the package, please
\usepackage[T1]{fontenc} % if needed

\usepackage{latexsym}
\usepackage{amssymb}
\usepackage{amsmath}
\usepackage{graphicx}
\usepackage{simplewick}
\usepackage{mathtools}
\usepackage{comment}
\usepackage{mleftright}
\usepackage{xparse}
\usepackage{physics}
\usepackage{multicol}
\usepackage{xcolor}

\def\one{\mbox{1 \kern-.59em {\rm l}}}

\def\half{\frac{1}{2}}

\def\C{\mathcal{C}}

\def\C{\mathcal{C}}

\def\F{\mathcal{F}}

\def\lr#1{\left(#1\right)}
\def\slr#1{\left[#1\right]}

\def\trl#1{\textrm{tr}\lr{#1}}

\def\avg#1{\left\langle #1\right\rangle}

\def\RS2{\mathbb R\times S^2_F}

\def \bal {\begin{align}}
\def \eal {\end{align}}

\definecolor{newBlue}{RGB}{94,129,181}
\definecolor{newRed}{RGB}{236,98,53}
\definecolor{newGreen}{RGB}{143,177,49}

\def \be  {\begin{equation}}
\def \ee  {\end{equation}}

\title{\boldmath Cubic asymmetric multitrace matrix model}

\author[a]{Benedek Bukor}
\author[a]{Juraj Tekel}

\affiliation[a]{Department of theoretical physics,\\Faculty of mathematics, physics and informatics, Comenius University in Bratislava,\\Mlynsk\'a Dolina, 8428 48, Bratislava, Slovakia}

\emailAdd{benedek.bukor@fmph.uniba.sk}
\emailAdd{juraj.tekel@fmph.uniba.sk}

\abstract{We analyze multitrace random matrix models with the help of the saddle point approximation and we introduce a multitrace term of type $-c_1c_3$ to the action. We obtain the numerical phase diagram of the model, with a stable asymmetric phase and the triple point. Furthermore, we examine response functions in this model.}


\begin{document} 
\maketitle
\flushbottom

\section{Introduction}\label{sec1}

Random matrix models provide us with a useful Swiss Army knife to numerous applications in theoretical physics. In nuclear physics, the distance between adjacent energy levels of heavy nuclei can be described by the difference between consecutive eigenvalues of random hermitian matrices. In particle physics, by the generalization of the ${3\times3}$ special unitary matrices of the gluon gauge field to infinite dimensional special unitary matrices, we arrive to random matrices due to the quantum mechanical uncertainty. In solid state physics, the growth of a crystal can be described by random matrices as well. For further applications see \cite{random}.

Turning to mathematical physics, an important application regards the quantum field theory defined on fuzzy spaces \cite{GKP,jtfuzzy}. The action of such a model is the sum of the kinetic term -- where the topology of the fuzzy space is encoded -- and the potential of the field. The pure quartic potential model is analytically solvable and well-examined, see \cite{random,introd,jt15b}. There are two realized phases of this model: the symmetric 1-cut and the symmetric 2-cut. 
In this study, we add a simple multitrace term to the potential in the form of $-c_1c_3$. The interpretation of the multitrace term can be twofold. Either it can be considered as an interaction term to the pure potential, or it can be considered as an approximate effective kinetic term of scalar field theory defined on a noncommutative space. Probability distributions in pure potential models are invariant under unitary transformations of the matrix, but the kinetic term in field theory breaks this invariance. To obtain an eigenvalue problem we need to integrate over the unitary -- or angular -- degrees of freedom and this integration leads to complicated multitrace contributions \cite{denjoe,jtfuzzy}.

On one hand, our choice of the $-c_1c_3$ multitrace term is motivated by \cite{ramda}, where the model was examined numerically by Monte Carlo simulations and the phase diagram has been obtained. In this case, the multitrace term described a type of interaction in the pure potential. The most significant impact of this term is that this addition introduces a stable asymmetric 1-cut phase. Existence of such solution is important, since such a phase has been observed in numerical analysis of fuzzy field theories, see e.g. \cite{Kovacik:2018thy,num1,num2,num3,panero15}.

On the other hand, this $-c_1c_3$  multitrace term appears in the effective action of the kinetic term of the Grosse-Wulkenhaar model. The Grosse-Wulkenhaar matrix model is a scalar field theory living on the truncated Heisenberg algebra. The two key features of this matrix model is its renormalizability and that the model is coupled to the curvature of the background of the noncommutative space. We will not dive into the mathematical analysis of the Grosse-Wulkenhaar model, this model gives us only a motivation, we refer the interested readers to \cite{GW1,trheal,dragan1,dragan2,svk2023} for further details.

In the presented work, we analyze the phase diagram of this multitrace matrix model. We demonstrate analytically that the model does indeed have a stable asymmetric phase and we identify the triple point in agreement with the previous numerical simulations. We also show that the model does not have a stable asymmetric 2-cut phase. We then study the response functions -- specific heat, magnetic susceptibility and string susceptibility -- and show that our approach can be successfully used to analyze also these more involved properties of the matrix models analytically. We point out agreement and discrepancy with the previous results. We also comment on previous results for the string susceptibility in a more simple symmetric multitrace model.

The problematic part of the model is in the asymmetric regime, where the conditions from the saddle point equation cannot be solved analytically. Our method employs a numerical solution of these equations, but we would like to stress that this is different from the numerical approach by Monte Carlo simulations.

The paper is structured as follows. In section \ref{sec2}, we present our preliminaries about the saddle point approximation, we review the results for the pure quartic potential and introduce our multitrace term. In section \ref{sec3}, we uncover the phase diagram of the model and find its triple point. In section \ref{sec4}, we investigate \textcolor{black}{response functions} of the model. 
 
\section{Review of matrix models}\label{sec2}
\subsection{A bird's-eye view of matrix models}\label{sec2.1}

This section gives a brief introduction into the concepts of matrix models that we will exploit, for further details see \cite{random,introd}. Let us start our discussion with the introduction of the ${N\times N}$ hermitian matrix models given by the integration Dyson measure $[\mathrm{d}M]$ and the probability measure $e^{-N^2S(M)}$. Then, the expectation value of function $f(M)$ is
\begin{align}
\avg{f}\,=\,\frac{1}{Z}\int [\mathrm{d}M] e^{-N^2 S(M)} f(M)\ \text{,}\quad \text{where}\quad Z\,=\,\int [\mathrm{d}M] e^{-N^2 S(M)}\ .\label{f}
\end{align}
Note that the action ${N^2S(M)}$ is of the same order as $[\mathrm{d}M]$.
The most general singletrace action is of the form
\begin{align}
S(M)\,=\, \frac{1}{N}\trl{V(M)}\ \text{,}\quad \text{where}\quad V(M)\,=\,\sum_{n=0}^{N}g_n\,M^n,\, g_n\in\mathbb{R}\ .\label{S}
\end{align}
Note that the factor $\frac{1}{N}$ ensures the correct order of the action when we take the trace of the potential\footnote{Note that the constant term in the definition of $V(M)$ can be eliminated because it will cancel thanks to \eqref{f}, anyway. We can get rid of the linear term as well by means of the redefinition of the matrix $M$. We will denote the coefficient in front of the quadratic term as $g_2=r/2$ and in front of the quadratic term as ${g_4=g}$.} $V(M)$. \textcolor{black}{If the action is invariant under the transformation ${M\to U M U^{\dag}}$, we can write ${M=U\Lambda U^{\dag}}$}, where $\Lambda $ is the diagonal matrix of the eigenvalues $\lambda_i$ of $M$ and ${U\in\text{U}(N)}$. 
The Jacobian of this transformation is
\begin{align}
[\mathrm{d}M]=[\mathrm{d}U]\lr{ \prod_{i=1}^N \mathrm{d}\lambda_i}\lr{ \prod_{i<j}\lr{\lambda_i-\lambda_j}^2}\ .
\end{align}
Provided that the function ${f(M)}$ is $U$-invariant, it depends only on the eigenvalues $\lambda_i$. The integration over the angular part in terms of the Haar measure $[\mathrm{d}U]$ is trivial, and the expectation value \eqref{f} becomes
\begin{align}
\avg{f}\,=\,\frac{1}{Z}\int \lr{ \prod_{i=1}^N \mathrm{d}\lambda_i} e^{-N^2 \slr{S(\Lambda)-\frac{2}{N^2}\sum\limits_{i<j}\text{ln}\lvert\lambda_i -\lambda_j\rvert}} f(\lambda_i )\ .\label{fn}
\end{align}
Let us define the quantity in the bracket in \eqref{fn} as the free energy (see \cite{introd})
\begin{align}
\mathcal{F}\,=\, S(\Lambda)-\frac{2}{N^2}\sum_{i<j}\text{ln}\lvert\lambda_i -\lambda_j\rvert\ .\label{FREE}
\end{align}
From now on, we will be interested exclusively in the case when ${N\rightarrow\infty}$. This means that all the contributions to the integral \eqref{fn} will be suppressed but that one stable configuration $\tilde\lambda_i$ %\textcolor{black}{of the matrix $\tilde M$}
which globally minimizes $\mathcal{F}$. \textcolor{black}{According to this consideration, we get the saddle point condition for discrete spectrum of the action \eqref{S} in the form}
\begin{align}
V'(\tilde\lambda_i)-\frac{2}{N}\sum_{j\neq i} \frac{1}{\tilde\lambda_i-\tilde\lambda_j}\,=\,0\ .\label{dspa}
\end{align}
%So, the expectation value straightforwardly becomes the function of the configuration $\tilde\lambda_i$
%\begin{align}
%\avg{f(\{\lambda_i\})}\longrightarrow f(\{\tilde\lambda_i\}) .\label{}
%\end{align}
Let us introduce for the discrete case the eingenvalue density $\rho(\lambda)$, the planar resolvent $\omega(z)$ and the moments of distribution $c_n$ in the way
\begin{align}
\rho(\lambda)\,=\,\frac{1}{N}\sum_{i=1}^N\delta\lr{\lambda-\tilde\lambda_i}\ \text{,}\,\, \,\,\omega(z)\,=\,\frac{1}{N}\sum_{i=1}^N\frac{1}{z-\tilde\lambda_i}\,\,\,\, \text{and}\,\,\,\, c_n\,=\,\frac{1}{N}\trl{M ^n}\ .\label{roc}
\end{align}
By the implementation of the limit ${N\rightarrow\infty}$, these functions turn to be continuous, thus
\begin{align}
\frac{1}{N}\sum_{i=1}^N f(\tilde \lambda_i)\longrightarrow \int_\C \mathrm{d}\lambda'\, \tilde\rho(\lambda')\, f(\lambda')\ ,\label{trans}
\end{align}
where the integral is over the support of the distribution $\C$. According to this transformation, we can rewrite the planar resolvent $\omega(z)$ and the moments of distribution $c_n$ in the way -- from now on we will omit tilde to refer to the solution of the saddle point equation --
\begin{align}
\omega(z)\,=\,\int_\C \mathrm{d}\lambda'\, \frac{\rho(\lambda')}{z-\lambda'}\,\,\,\, \text{and}\,\,\,\, c_n=\int_\C \mathrm{d}\lambda'\,\rho(\lambda')\lambda'^n\ .\label{oc}
\end{align}
In light of \eqref{trans}, the saddle point equation \eqref{dspa} transforms to 
\begin{align}
V'(\lambda)-2\, \text{PV}\int_\C \mathrm{d}\lambda'\, \frac{\rho(\lambda')}{\lambda-\lambda'}\,=\,0\ ,\label{vspe}
\end{align}
where PV denotes the principal value of the integral. By means of the Sokhotski-Plemelj formula,
\begin{align}
\text{PV}\int_\C \mathrm{d}\lambda'\, \frac{\rho(\lambda')}{\lambda-\lambda'}\,=\,\lim_{\varepsilon\rightarrow0^+}\slr{\omega(\lambda\pm\text{i}\varepsilon)\pm\text{i}\pi\rho(\lambda)}\ ,\label{}
\end{align}
we can rewrite \eqref{vspe} to the form
\begin{align}
V'(\lambda)\,=\,\lim_{\varepsilon\rightarrow0^+}\slr{\omega(\lambda+\text{i}\varepsilon)+\omega(\lambda-\text{i}\varepsilon)}\ ,\label{}
\end{align}
and get $\rho(\lambda)$ as
\begin{align}
\rho(\lambda)\,=\,\frac{1}{2\pi\text{i}}\,\lim_{\varepsilon\rightarrow0^+}\slr{\omega(\lambda+\text{i}\varepsilon)-\omega(\lambda-\text{i}\varepsilon)}\ .\label{}
\end{align}
Computing the square of the resolvent \eqref{roc}, neglecting all the subdominant terms and taking the advantage of the saddle point equation \eqref{dspa}, we get a quadratic equation whose solution is 
\begin{align}
%\omega(\lambda)^2\,=\,V'(\lambda)\,\omega(\lambda)-P(\lambda)\ ,\label{}
\omega(z)=\frac12\slr{V'(z)-\sqrt{V'(z)^2-4\,P(z)}}\ ,\label{omega1}
\end{align}
where $P(z)$ is a polynomial of a specific uninteresting form. The expression under the square root splits into two parts -- an  even polynomial $\sigma(z)$ which determines the support of the distribution, and the rest -- which is positive, analytic and does not contribute to the discontinuity denoted as $\lvert H(z)\rvert$. We can thus write \eqref{omega1} as
\begin{align}
\omega(z)=\frac12\slr{V'(z)-\lvert H(z)\rvert\sqrt{\sigma(z)}}\ .\label{omega2} 
\end{align}
We will, in general, tackle two cases\footnote{Theoretically, assumption with more cuts can be tackled, as well. However, we will see that the choice of our action \eqref{Sm} does not allow these possibilities.}
\begin{itemize}
  \item the 1-cut solution
  \begin{align}
  \sigma(\lambda)&=\slr{\lambda-\lr{D-\sqrt{\delta}}}\slr{\lambda-\lr{D+\sqrt{\delta}}}\nonumber\\
  \C&=(\,D-\sqrt{\delta}\, ,\,D+\sqrt{\delta}\,)\, ,
  \end{align}
  \item the 2-cut solution\\
  \begin{align}
  \sigma(\lambda)&=\slr{\lambda+\sqrt{D_1+\delta_1}}\slr{\lambda+\sqrt{D_1-\delta_1}}\slr{\lambda-\sqrt{D_2-\delta_2}}\slr{\lambda-\sqrt{D_2+\delta_2}}\nonumber\\
  \C&=(\,-\sqrt{D_1+\delta_1}\, ,\,-\sqrt{D_1-\delta_1}\,)\,\cup\,(\,\sqrt{D_2-\delta_2}\, ,\,\sqrt{D_2+\delta_2}\,)\, . \label{Comega2}
  \end{align} 
\end{itemize}
The function $H(z)$ is determined by the requirement obtained from \eqref{oc},
\begin{align}
\omega(z)=\frac{1}{z}\,\,\,\,\text{for}\,\,\,\,\lvert z\rvert\rightarrow\infty\ .\label{omega3}
\end{align}
According to this condition, the polynomial part of \eqref{omega2} in ${\lvert z\rvert\rightarrow\infty}$ must be $0$, which determines the function $\lvert H(z)\rvert$ and the endpoints of the support $\C$.
Eventually, utilizing \eqref{omega2}, we get 
\begin{align}
\rho(\lambda)=\frac{1}{2\text{i}\pi}\lvert H(\lambda)\rvert\sqrt{\sigma(\lambda)}\ .\label{} 
\end{align}
Now, we conclude that the expectation value becomes
%\footnote{Thereby, the  eigenvalue density $\rho(\lambda)$ is the distribution function.}
\begin{align}
\avg{f}\,=\,\int_\C \mathrm{d}\lambda\, \rho(\lambda)\, f(\lambda)\ .
\end{align}
The last remarkable trait of this derivation, which is definitely worth mentioning, is that the expansion of the planar resolvent \eqref{oc}  yields the moments of the distribution $c_n$, 
\begin{align}
\omega(z)\,=\,\frac{1}{z}\int_\C \mathrm{d}\lambda'\, \sum_{n=0}^{\infty}\lr{\frac{\lambda'}{z}}^n\rho(\lambda')\,=\,\sum_{n=0}^{\infty}\frac{c_n}{z^{n+1}}\ .\label{omega4}
\end{align}

\subsubsection*{Free energy}\label{}

Let us turn back to the definition of the free energy \eqref{FREE}, which in the continuous case transforms to 
\begin{align}
\mathcal{F}\,=\, \int_\C \mathrm{d}\lambda\, \rho(\lambda)\, V(\lambda)-\iint_{\C\times\C\setminus \{\lambda=\tau\}}\mathrm{d}\lambda\mathrm{d}\tau \rho(\lambda) \rho(\tau)\ln\lvert\lambda -\tau\rvert\ .\label{FREE1}
\end{align}
The double integral is cumbersome to calculate but there is an ingenious trick to handle it. \\

Let us first describe it in the case of the 1-cut solutions. We would like to remind the reader that the free energy was minimized in \eqref{dspa}, and the trick is that it will be minimized again simultaneously with the Lagrange multiplier $\xi$ which ensures the proper normalization of the eigenvalue density $\rho(\lambda)$. So, we should minimize the following action for the 1-cut solution
\begin{align}
S_{V1}(\rho)\,=\, \mathcal{F}+\xi\lr{1-\int_\C \mathrm{d}\lambda\, \rho(\lambda)}\ \label{}
\end{align}
by the variation with respect to $\rho(\lambda)$, where the subscript signifies the difference of this action from the original one \eqref{S}. After the variation, we get the Lagrange multiplier for the 1-cut solution
\begin{align}
\xi\,=\, V(\lambda)-2\int_{\C}\mathrm{d}\tau  \rho(\tau)\ln\lvert\lambda -\tau\rvert\ ,\label{xi}
\end{align}
which is the same for every ${\tau\in\C}$. Substituting the integral from \eqref{xi} to \eqref{FREE1}, we obtain the free energy for the 1-cut case.
\begin{align}
\mathcal{F}_1\,=\,\frac12\lr{\int_\C \mathrm{d}\lambda\, \rho(\lambda)\, V(\lambda) +\xi}\ .\label{F}
\end{align}
 
The presented trick can be executed for the 2-cut case as well, but with a slight twist which has not been described so far based on the knowledge of the authors. Let us denote that part of the eigenvalue density $\rho(\lambda)$ which is supported on the interval ${\C_1=(\,-\sqrt{D_1+\delta_1}\, ,\,-\sqrt{D_1-\delta_1}\,)}$ as $\rho_1(\lambda)$ and the part which is supported on the interval ${\C_2=(\,\sqrt{D_2-\delta_2}\, ,\,\sqrt{D_2+\delta_2}\,)}$ as $\rho_2(\lambda)$. The difference from the 1-cut case is that we do not utilize the normalization but for a given eigenvalue density we fix the so-called filling fraction which gives the ratio of the "amount" of the eigenvalues found in a given support to the "whole amount". So, we should minimize the following action for the 2-cut solution
\begin{align}
S_{V2}(\rho)\,=\, \mathcal{F}+\xi_1\lr{f_1-\int_{\C_1} \mathrm{d}\lambda\, \rho_1(\lambda)}+\xi_2\lr{f_2-\int_{\C_2} \mathrm{d}\lambda\, \rho_1(\lambda)}\ \label{}
\end{align}
by the variation with respect to $\rho_1(\lambda)$ and $\rho_2(\lambda)$. After the variation, we get the Lagrange multipliers for the 2-cut solution
\begin{align}
\xi_1\,&=\, V(\lambda)-2\slr{\int_{{\C_1}}\mathrm{d}\tau  \rho_1(\tau)\ln\lvert\lambda -\tau\rvert+\int_{{\C_2}}\mathrm{d}\tau  \rho_2(\tau)\ln\lvert\lambda -\tau\rvert}\,\, \text{for}\,\, \lambda\in\C_1\,\, ,\nonumber \\
\xi_2\,&=\, V(\lambda)-2\slr{\int_{{\C_1}}\mathrm{d}\tau  \rho_1(\tau)\ln\lvert\lambda -\tau\rvert+\int_{{\C_2}}\mathrm{d}\tau  \rho_2(\tau)\ln\lvert\lambda -\tau\rvert}\,\, \text{for}\,\, \lambda\in\C_2\,\, .
\label{xi12}
\end{align}
Substituting the proper integrals from \eqref{xi12} to \eqref{FREE1}, we obtain the free energy for the 2-cut case.
\begin{align}
\mathcal{F}_2\,=\,\frac12\lr{\int_\C \mathrm{d}\lambda\, \rho(\lambda)\, V(\lambda) +\xi_1\int_{{\C_1}}\mathrm{d}\lambda\, \rho_1(\lambda)+\xi_2\int_{{\C_2}}\mathrm{d}\lambda\, \rho_2(\lambda)}\ .\label{F2}
\end{align}
  
\subsection{Results for the quartic potential}\label{sec2.2}
First, we will summarize the results for the quartic potential with the action
\begin{align}
S(M)\,=\,\frac{1}{N}\slr{\frac12r\,\trl{M^2}+g\,\trl{M^4}}\ \text{,}\quad g\in\mathbb{R}^+\ .\label{Sm}
\end{align}
The motivation for this potential is twofold. This is the least convoluted potential possible, and corresponds to the quantum scalar field theory with interaction of type $g\phi^4$ on fuzzy spaces.

We  will concentrate only on the case of symmetric cuts. The exact analysis of this action can be found in \cite{jt15b}. With  symmetric 1-cut assumption we get
%        \leavevmode\vspace*{-\dimexpr\abovedisplayskip + \baselineskip}
\begin{align}\label{s1}
\quad\text{support}\quad &\C_{\text{S1}}=(\,-\sqrt{\delta}\, ,\,\sqrt{\delta}\,)\,\text{, where}\,\,\, \delta\,=\,\frac{1}{6g}\lr{\sqrt{r^2+48g}-r}\ ,\nonumber\\
\quad\text{eigenvalue density}\quad&\rho_{\text{S1}}(\lambda)\,=\,\frac{1}{2\pi}\lr{r+2g\delta+4g\lambda^2}\sqrt{\delta-\lambda^2}\ ,\nonumber\\      
\quad\text{free energy}\quad &\F_{\text{S1}}\,=\,\frac{-r^2\delta^2+40r\delta}{384}-\frac12\ln{\frac{\delta}{4}}+\frac38\ 
,\\
\quad\text{first/third moment}\quad &\lr{c_1}_{\text{S1}}\,=\,\lr{c_3}_{\text{S1}}\,=\,0\ .\nonumber
\end{align}
This solution however becomes negative for for ${r<-4\sqrt{g}}$ and is no longer a well defined eigenvalue density. We thus need to make the the symmetric 2-cut assumption, which yields
\begin{align}\label{s2}
\quad\text{support}\quad &\C_{\text{S2}}=(\,-\sqrt{D+\delta}\, ,\,-\sqrt{D-\delta}\,)\,\cup\,(\,\sqrt{D-\delta}\, ,\,\sqrt{D+\delta}\,)\ ,\nonumber\\ 
&\text{where}\,\,\, \delta\,=\,\sqrt{\frac{1}{g}} \quad \text{and}\quad D\,=\,-\frac{r}{4g}\ ,\nonumber\\
\quad\text{eigenvalue density}\quad&\rho_{\text{S2}}(\lambda)\,=\,\frac{2g\lvert\lambda\rvert}{\pi}\sqrt{\delta^2-\lr{D-\lambda^2}^2}\ ,\nonumber \\      
\quad\text{free energy}\quad &\F_{\text{S2}}\,=\,-\frac{r^2}{16g}+\frac14\ln{4g}+\frac38\ ,\\
\quad\text{first/third moment}\quad &\lr{c_1}_{\text{S2}}\,=\,\lr{c_3}_{\text{S2}}\,=\,0\ .\nonumber
\end{align}
It is worth mentioning that for ${r<-2\sqrt{15g}}$ there exists a third solution, an asymmetric 1-cut solution, which is, however, not realized since its free energy is greater than the free energy of the symmetric 2-cut case. Nevertheless, as we will see shortly, it will play a crucial role in the model with the multitrace term.

\subsection{Introduction of the multitrace term $c_1c_3$}\label{sec2.3}
As opposed to the singletrace matrix models, e.g. \eqref{Sm}, the action of the multitrace matrix models also involves the product of the trace of some powers of the matrix.
We will focus on the multitrace term type of $c_1c_3$, which modifies the action to
\begin{align}
S(M)\,=\,\frac{1}{N}\slr{\frac12r\,\trl{M^2}+g\,\trl{M^4}}+t\,c_1c_3\,\text{,}\,\, \,\,t\in\mathbb{R}\ ,\label{multiSm}
\end{align}
with the free energy 
\begin{align}
\mathcal{F}\,=\,\frac{1}{N}\slr{\frac12r\,\trl{\Lambda^2}+g\,\trl{\Lambda^4}}+tc_1c_3 -\frac{2}{N^2}\sum_{i<j}\text{ln}\lvert\lambda_i -\lambda_j\rvert\ .\label{FREEm}
\end{align}
For this action \eqref{multiSm}, we get the saddle point equation \eqref{dspa} of the form
\begin{align}
\lr{\frac12 2r\tilde\lambda_i+4g\tilde\lambda_i^3+3tc_1\tilde\lambda_i^2+tc_3}-\frac{2}{N}\sum_{j\neq i} \frac{1}{\tilde\lambda_i-\tilde\lambda_j}\,=\,0\ .\label{spm}
\end{align}

However, in the large $N$ limit a singletrace matrix model with the effective potential
\begin{align}
V_{\text{eff}}(M)\,=\,\frac12rM^2+gM^4+tc_1M^3+tc_3M\ \label{Veff}
\end{align}
would yield the exact same saddle point condition as \eqref{spm}.
For this effective potential \eqref{Veff} the effective free energy is
\begin{align}
\mathcal{F}_{\text{eff}}\,=\,\frac{1}{N}\slr{\frac12r\,\trl{\Lambda^2}+g\,\trl{\Lambda^4}}+2t\,c_1c_3 -\frac{2}{N^2}\sum_{i<j}\text{ln}\lvert\lambda_i -\lambda_j\rvert\ .\label{FREEe}
\end{align}
By the subtraction of \eqref{FREEm} and \eqref{FREEe}, we get that the free energy of our multitrace model \eqref{multiSm} is
\begin{align}
\mathcal{F}\,=\,\mathcal{F}_{\text{eff}}-tc_1c_3\ .\label{Feff}
\end{align}

In order to calculate the effective free energy $\mathcal{F}_{\text{eff}}$, the effective model \eqref{Veff} could be solved by the standard methods described in the previous section or in the literature \cite{random,jt15b}. The complication is that the parameters of the effective model now depend on the moments of the distribution, which themselves depend on the parameters of the model. Thus equations for $c_1$ and $c_3$ coming from the resolvent \eqref{omega4} are not independent and must be solved together with the equations defining the endpoints of the cut(s). This consideration leads to a set of self-consistent equations.

After all the dust settles, we obtain the following set of equations for the asymmetric 1-cut solution with the support ${(\,D-\sqrt{\delta}\, ,\,D+\sqrt{\delta}\,)}$;
\begin{align}
    0\,=\,&2 D^3 g + 3 D \delta g +\frac{1}{2} D r + \lr{\half c_3 + \frac{3}{2} c_1 D^2 + \frac{3}{4}c_1 \delta} t\ ,\nonumber\\
    1\,=\,&3 D^2 \delta g + \frac{3}{4}\delta^2 g+\frac{1}{4}\delta r+\frac{3}{2}c_1 D \delta t\ ,\nonumber\\
    c_1\,=\,&3 D^3 \delta g+\frac{3}{2}D \delta^2 g +\frac{1}{4}D \delta r+\lr{\frac{3}{2}c_1 D^2 \delta  + \frac{3}{16}c_1 \delta^2} t\ ,\nonumber\\
    c_3\,=\,&3 D^5 \delta g+\frac{21}{4} D^3 \delta^2 g+\frac{9}{8} D \delta^3 g+\frac{1}{4} D^3 \delta r + \frac{3}{16} D \delta^2 r +\nonumber\\&+\lr{\frac{3}{2} c_1 D^4 \delta  + \frac{27}{16} c_1 D^2 \delta^2  + \frac{3}{32} c_1 \delta^3} t\ .
\end{align}
The set of equations for the asymmetric 2-cut solution with the support\footnote{Note that this choice of the support is different from \eqref{Comega2}. Equations for the asymmetric 2-cut solution in this choice become much more tractable.} \[\lr{\,-\lr{D_1+\sqrt{\delta_1}}\, ,\,-\lr{D_1-\sqrt{\delta_1}}\,}\,\cup\,\lr{\,D_2-\sqrt{\delta_2}\, ,\,D_2+\sqrt{\delta_2}\,}\]
is 
\begin{align}
    0\,=\,&\frac32 c_1\lr{ D_1 + D_2 }t+\lr{\delta_1 +\delta_2} g+2\lr{ D_1^2 +D_1 D_2 + D_2^2 }g+\frac{r}{2}\ ,\nonumber\\
    0\,=\,&\lr{\frac{3 c_1 \delta_1 }{4}+\frac{3 c_1 \delta_2 }{4}-\frac{3}{2} c_1 D_1 D_2 +\frac{c_3 }{2}}t+2 \lr{\delta_1 D_1 + \delta_2 D_2} g-\nonumber \\
    &-2 D_1 D_2 \lr{ D_1 +D_2} g\ ,\nonumber\\
    1\,=\,&\frac{3}{4}c_1 t\lr{\delta_1-\delta_2}\lr{D_1-D_2}+\frac14\lr{\delta_1-\delta_2}^2g+2 \delta_1 D_1^2 g+2 \delta_2 D_2^2 g-\nonumber \\
    &-\delta_1 D_1 D_2 g-\delta_1 D_2^2 g-\delta_2D_1^2  g-\delta_2 D_1  D_2 g\ ,\nonumber\\
    c_1\,=\,&\frac{3}{16} c_1 \delta_1^2 t+\frac{3}{4} c_1 \delta_1 D_1^2 t-\frac{3}{4} c_1 \delta_1 D_1 D_2 t-\frac{3}{8} c_1 \delta_1 \delta_2 t-\frac{3}{4} c_1 \delta_2 D_1 D_2 t+\nonumber \\
    &+\frac{3}{16} c_1 \delta_2^2 t+\frac{3}{4} c_1 \delta_2 D_2^2 t+\delta_1^2 D_1 g+2 \delta_1 D_1^3 g-\delta_1 D_1^2 D_2 g-\delta_1 \delta_2 D_1  g-\nonumber \\
    &-\delta_1 D_1 D_2^2 g-\delta_1 \delta_2 D_2 g- \delta_2 D_1^2  D_2 g-\delta_2 D_1 D_2^2 g+\delta_2^2 D_2 g+2 \delta_2 D_2^3 g\ ,\nonumber\\
    c_3\,=\,&\frac{3}{32} c_1 \delta_1^3 t+\frac{9}{8} c_1 \delta_1^2 D_1^2 t-\frac{9}{16} c_1 \delta_1^2 D_1 D_2 t-\frac{3}{32} c_1 \delta_1^2 \delta_2 t+\frac{3}{4} c_1 \delta_1 D_1^4 t-\nonumber \\
    &-\frac{3}{4} c_1 \delta_1 D_1^3 D_2 t-\frac{3}{8} c_1 \delta_1 D_1^2 \delta_2 t-\frac{3}{8} c_1 \delta_1 D_1 \delta_2 D_2 t-\frac{3}{32} c_1 \delta_1 \delta_2^2 t-\nonumber \\
    &-\frac{3}{8} c_1 \delta_1 \delta_2D_2^2 t-\frac{9}{16} c_1 D_1 \delta_2^2 D_2 t-\frac{3}{4} c_1 D_1 \delta_2 D_2^3 t+\frac{3}{32} c_1 \delta_2^3 t+\nonumber \\
    &+\frac{9}{8} c_1 \delta_2^2 D_2^2 t+\frac{3}{4} c_1\delta_2 D_2^4 t+\frac{3}{4} \delta_1^3 D_1 g+4 \delta_1^2 D_1^3 g-\frac{3}{4} \delta_1^2 D_1^2 D_2 g-\nonumber \\
    &-\frac{1}{2} \delta_1^2 D_1 \delta_2 g-\frac{3}{4} \delta_1^2 D_1 D_2^2 g-\frac{1}{4} \delta_1^2 \delta_2 D_2 g+2 \delta_1 D_1^5 g-\delta_1 D_1^4 D_2 g-\nonumber \\
    &-\delta_1 D_1^3\delta_2 g-\delta_1 D_1^3 D_2^2 g-\frac{3}{2} \delta_1 D_1^2 \delta_2 D_2 g-\frac{1}{4} \delta_1 D_1 \delta_2^2 g-\frac{3}{2} \delta_1 D_1\delta_2 D_2^2 g-\nonumber \\
    &-\frac{1}{2} \delta_1 \delta_2^2 D_2 g-\delta_1 \delta_2D_2^3 g-\frac{3}{4} D_1^2 \delta_2^2 D_2 g-D_1^2 \delta_2 D_2^3 g-\frac{3}{4} D_1\delta_2^2D_2^2 g-\nonumber \\
    &-D_1\delta_2 D_2^4 g+\frac{3}{4} \delta_2^3 D_2 g+4\delta_2^2 D_2^3 g+2\delta_2 D_2^5 g\ .
\end{align}
This illustrates what we mean by the self-consistent conditions. 

\section{The phase diagram}\label{sec3}

The first property of the model that we will look at is the phase diagram. This means that we will look for the solution, or perhaps solutions, of the saddle point equation for given values of parameters and try to locate the transition lines in between the regions of given topology of the support of the distribution with the lowest free energy.

The acquired equations of our model -- some of them shown in the previous section -- cannot be solved analytically. To obtain the phase diagram and other properties of the model, we have to solve them numerically.

\subsection{Description of the numerical calculations}\label{desc}
The process in which we obtained the phase diagram goes as follows. The ingredients, which are imposed by hand, are 
\begin{itemize}
  \item the potential with(out) multitrace terms: the action for the initial pure potential is \eqref{Sm}, and the action of the investigated multitrace term is \eqref{multiSm},
  \item either the 1-cut or the 2-cut assumption. 
\end{itemize}
For each point $(g,\,r)$ of the phase space we searched for the 1-cut and the 2-cut solution numerically. Although, the quartic potential \eqref{Sm} was treated analytically, generally, the treatment of more complex actions and the asymmetric solutions is more burdensome, and one is left only with the numerical approach.

The 1-cut distribution function for a given $(g,\,r)$ can be obtained  following the instructions as described in \ref{sec2.1}. Using \eqref{omega2} and  \eqref{omega3}, we find all the possible $\lvert H(\lambda)\rvert$, which must be positive functions on their support. Suppose we had multiple positive 1-cut distributions, we would opt for the one with the lowest free energy.
In the case of pure potential \eqref{Sm}, the function $\lvert H(\lambda)\rvert$ is quadratic, so with the endpoints we have 3+2 parameters to be fixed. Since we have 5 constraints coming from the powers of $\lambda$, these can be fixed completely.

The search for the 2-cut solution calls for an analogous treatment as for the case of the 1-cut solution. Nevertheless, as opposed to the 1-cut case there is a considerable difference. In the case of pure potential \eqref{Sm}, the function $\lvert H(\lambda)\rvert$ 
%-- which should be positive by default -- 
is linear, so with the endpoints we have 2+4 parameters to be fixed. However, since we have 5 constraints coming from the powers of $\lambda$, these can not be fixed completely. One way around this issue is as follows. We treat one of the parameters as a free parameter and vary it. The support of the asymmetric 2-cut solution looks like ${(\,-\sqrt{D_1+\delta_1}\, ,\,-\sqrt{D_1-\delta_1}\,)\,\cup\,(\,\sqrt{D_2-\delta_2}\, ,\,\sqrt{D_2+\delta_2}\,)}$, and for instance we can scrutinize all the possible solutions for different ${k\in\mathbb{R}}$ when ${\delta_2=k\,\delta_1}$. Suppose that ${\delta_1\geq\delta_2}$ so ${k\in[0,1]}$. We slice up this interval, and we probe whether a solution exists with the currently examined endpoints of the support. Slicing up the interval sufficiently finely, we can get a sufficient number of possible solutions to opt for the one with the lowest free energy. The finer we chop, the more self-assured we can be that solution we found is getting closer to the real one and this controls our error.

We have tested our numerical calculations for the pure potential \eqref{Sm}, and we recovered the exact phase diagram as it had been analytically calculated in section \ref{sec2.2}, which confirmed the correctness of our numerical treatment. 
%The introduction of the multitrace term $c_1c_3$ brings further complications. We embark on the analysis again using the equations \eqref{omega2} and  \eqref{omega3}. But, on the one hand, the moments appear in the effective potential \eqref{Veff} of the action \eqref{multiSm}. On the other hand, they emerge by virtue of the expansion of $\omega(z)$ in \eqref{omega4}. To make it clear, the moments we acquire after the planar resolvent is calculated are in the input as well. This consideration leads to self-consistent equations in the numerical treatment, whose solution modifies the eigenvalue distribution. The calcualtion of the free energy $\mathcal{F}$ for the solution is described in \ref{sec2.3}.

\subsection{Phase diagram of the multitrace term $tc_1c_3$}\label{phasediag}

In the light of the previous section, we generated the phase diagram of the action \eqref{multiSm} for general 1-cut and general 2-cut solution in the case of four different ${t\in\{-0.5;-1;-2;-5\}}$ and thoroughly examined the solutions. As opposed to to the pure potential, we observed not only symmetric eigenvalue distributions as stable solutions, but for a specific region of the phase space the asymmetric 1-cut solution had the lowest free energy. We did not identify any points where a general asymmetric 2-cut solution would have lower free energy than the corresponding asymmetric 1-cut solution.
%Whereas the multitrace term on a certain part of the symmetric phases had no effect, on an other part of the phase space stabilized the asymmetric 1-cut phase. %\textcolor{red}{JURO In the vicinity of the transition line between the asymmetric 1-cut and symmetric 2-cut phases, we observed for some points asymmetric 2-cut solutions which in neither case were stable.}
%Thereby, for the intact regions of the symmetric phases the analytical solutions from section \ref{sec2.2} became handy. 
%So, at this point, we stopped searching for a general solution in each point $(g,\,r)$. 
As the multitrace terms do not change the symmetric regime of the model, the symmetric phases were right away obtained from the analytical approach using the formulae from section \ref{sec2.2}, the asymmetric 1-cut solution was still found by the numerical treatment. Let us stress that this refers to numerical solution of analytic equations and not to numerical simulation of the corresponding matrix integrals.
The free energy coming from the symmetric 1-cut or 2-cut solutions was compared with the free energy coming from the asymmetric 1-cut solution, 
and the phase whose free energy was the lowest was announced the stable one.

\subsubsection*{Phase diagram for ${t=-1}$}\label{}

\begin{figure}[h]
    \centering
    \includegraphics[width=\linewidth]{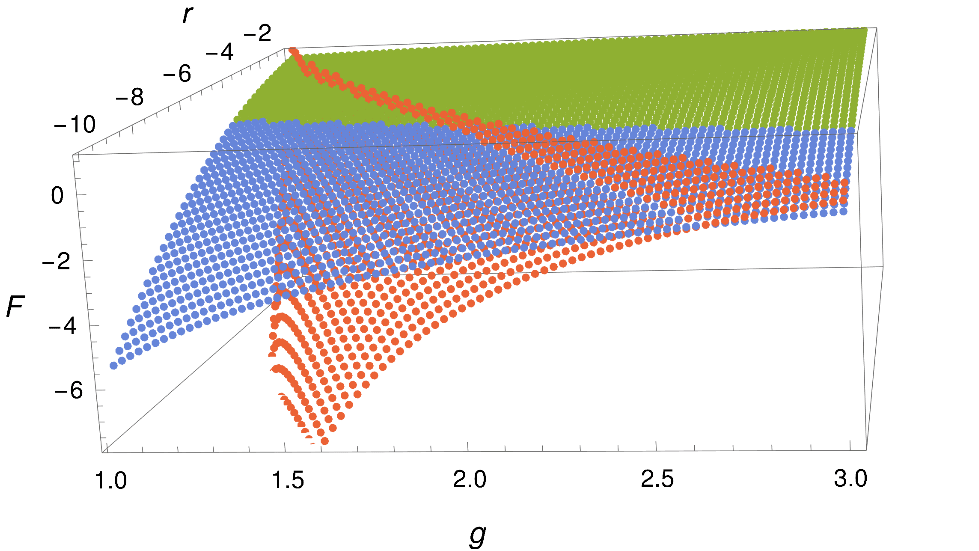}
    \caption{The free energies of the three phases of the model \eqref{c1c3}.\\
    The symmetric 1-cut solution is denoted with the \textcolor{newGreen}{green colour}, the symmetric 2-cut solution is denoted with the \textcolor{newBlue}{blue colour} and the asymmetric 1-cut solution is denoted with the \textcolor{newRed}{red colour}.}
    \label{free0}
\end{figure}

For the first case let us thoroughly investigate the action \eqref{multiSm} with the parameter ${t=-1}$, so we get
\begin{align}
S(M)\,=\,\frac{1}{N}\slr{\frac12r\,\trl{M^2}+g\,\trl{M^4}}-c_1c_3\ .\label{c1c3}
\end{align}
The correspondent comparison of the free energies illustrates Fig. \ref{free0}.

\begin{figure}[h]
    \centering
    \includegraphics[width=0.91\linewidth]{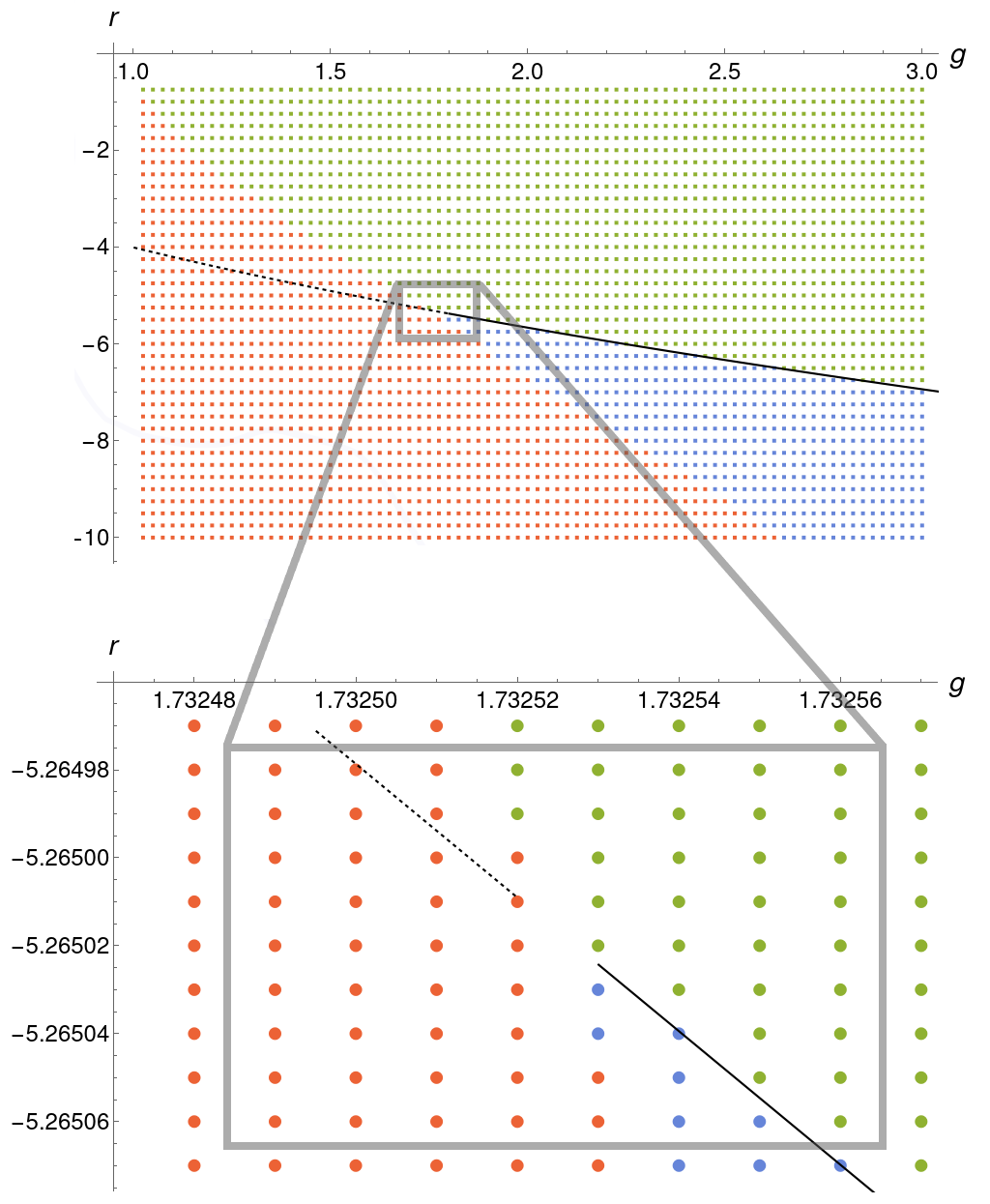}
    \caption{The phase diagram of the quartic potential with multitrace model \eqref{c1c3}.\\
  \textcolor{black}{The notation is  the same as in the case of Fig. \ref{free0}.} The black line denotes the phase transition line between the symmetric 1-cut and 2-cut solutions given by the equation ${r=-4\sqrt{g}}$, the dashed part denotes the region where the symmetric phase is not stable, while the continuous part is the real transition line between the symmetric 1-cut and symmetric 2-cut. \\
  Note that the lower gray rectangle is not the proportionally scaled version of the upper gray rectangle. Our intention is to mark out the region to be magnified in the upper diagram. For proper understanding, take into consideration the numbers on the axes.}
    \label{phase0}
\end{figure}

The obtained phase diagram of the action \eqref{c1c3} is Fig. \ref{phase0}.

Note that the region $g\leq 1$ is inaccessible for the asymetric regime of the action \eqref{c1c3}. It is a general feature of the actions with negative multitrace terms that in certain part of the parameter space the model does not have any solutions. In this particular region beyond our reach\footnote{Before releasing our numerical calculations to \eqref{c1c3}, we have checked the correctness of our numerical method on the action 
\begin{align*}
S(M)\,=\,\,\frac{1}{N}\slr{\frac12r\,\text{tr}(M^2)+g\,\text{tr}(M^4)}\pm c_2^2\ .\label{} 
\end{align*}
In both of the cases, we got utter agreement with \cite{jt15b, rus}. To confirm the traits of the actions with negative multitrace terms, in the case of multitrace term type of $-c_2^2$, and in the region ${g\leq 1,\, r\leq 0}$, we found no solution. 
}, only the symmetric solutions of the pure potential \eqref{Sm} are present. Here, the model has no stable asymmetric solution. This means that in this part of parameter space our simple model would not be a good approximation to realistic models, such as ones describing fuzzy field theories.

\subsubsection*{Triple point for ${t=-1}$}\label{}

Zooming in on the region where the three phases meet, we have a closer look at the triple point $(g_c,\,r_c)$, i.e. that point of the phase space where all the three phases coexist. By looking at the intersection of the phase transition lines between to adjacent phases, the triple point can be, in principle, found. However, we have no analytic knowledge about the function of the transition line which separates the asymmetric 1-cut solution from the symmetric ones.
%The obtained phase diagram of the action \eqref{c1c3} in the vicinity of the triple point is Fig. \ref{phase3}.

%\begin{figure}[h]
%    \centering
%    \includegraphics[width=\linewidth]{phase3.png}
%    \caption{The phase diagram of the quartic potential with multitrace model \eqref{c1c3} in the vicinity of the triple point.\\
%    The notation is  the same as in the case of Fig. \ref{free0}.}
%    \label{phase3}
%\end{figure}

In light of the acquired phase diagram, the coordinates of the triple point $(g_c,\,r_c)$ can be easily read out by following the colors of the dots,
\begin{align}
g_c\,=\, +1.732525\pm0.000005\quad \text{and}\quad r_c\,=\, -5.26502\pm0.00001\ .\label{triple}
\end{align}

The uncertainty of the coordinates of the triple point comes from the sampling step in parameters in Fig. \ref{phase0}.
The same value is obtained by using interpolating functions to extrapolate the numerical data in Fig. \ref{free0} into regions between the obtained values and looking numerically for intersection of surfaces given by such functions.

This exact same model has been investigated in \cite{ramda} by means of Monte Carlo simulations. There, the coordinates of the triple point in our notation are ${(1.7706\,;\,-5.5432)}$, whereas we obtained the triple point ${(g_c, r_c)\doteq(1.7325, -5.2650)}$. The difference is due to the fact that we tackle the problem in the strict large $N$ limit, while the Monte Carlo simulations have been done for finite sizes of matrices.

\newpage

\subsubsection*{Phase diagram and triple point for ${t=-0.5}$}\label{}

The action \eqref{multiSm} with the parameter ${t=-0.5}$ is
\begin{align}
S(M)\,=\,\,\frac{1}{N}\slr{\frac12r\,\trl{M^2}+g\,\trl{M^4}}-\frac12c_1c_3\ ,\label{c1c30.5}
\end{align}
\vspace{-1cm}
\begin{figure}[h]
\minipage{0.49\textwidth}
    \centering
    \includegraphics[width=\linewidth]{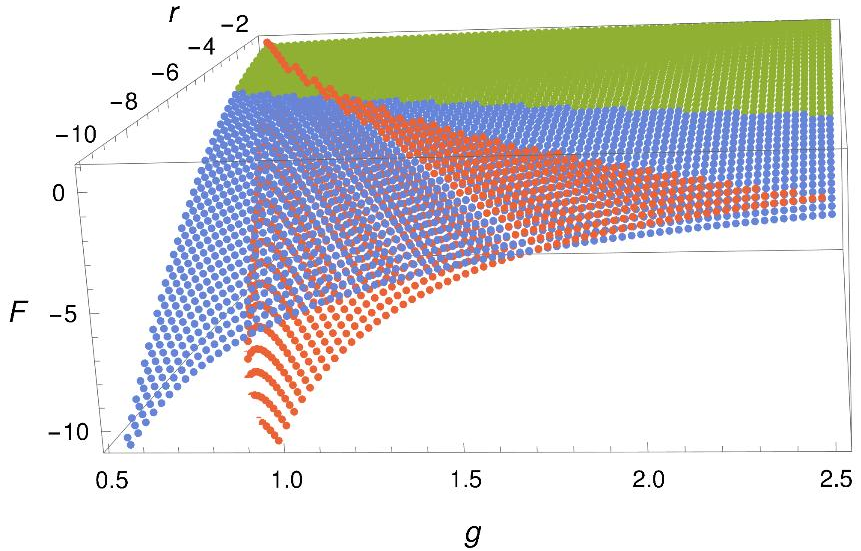}
    \caption{The free energies of the three phases of the model \eqref{c1c30.5} with the notation from Fig. \ref{free0}.}
    \label{free0.5}
\endminipage\hfill
\minipage{0.48\textwidth}

The correspondent comparison of the free energies illustrates Fig. \ref{free0.5}. This time the region $g\leq 0.5$ is inaccessible for this action \eqref{c1c30.5}.\\
The coordinates of the triple point are
\begin{align}
g_c\,&=\, +0.866265\pm0.000005\nonumber\\
r_c\,&=\, -3.722935\pm0.000015\ .
\end{align}

\endminipage\hfill
\end{figure}

\subsubsection*{Phase diagram and triple point for ${t=-2}$}\label{}

The action \eqref{multiSm} with the parameter ${t=-2}$ is
\begin{align}
S(M)\,=\,\,\frac{1}{N}\slr{\frac12r\,\trl{M^2}+g\,\trl{M^4}}-2c_1c_3\ .\label{c1c32.0}
\end{align}
\vspace{-1cm}
\begin{figure}[h]
\minipage{0.49\textwidth}
    \centering
    \includegraphics[width=\linewidth]{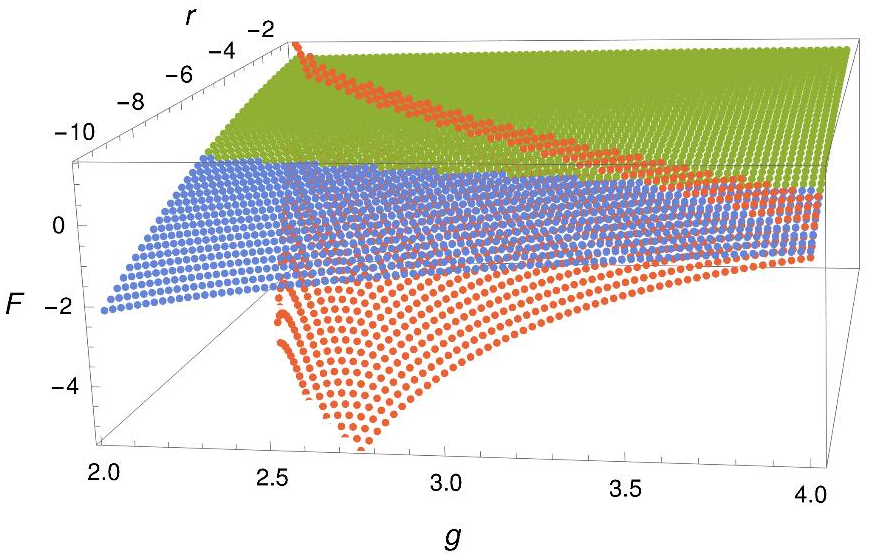}
    \caption{The free energies of the three phases of the model \eqref{c1c32.0} with the notation from Fig. \ref{free0}.}
    \label{free2.0}
\endminipage\hfill
\minipage{0.48\textwidth}

The correspondent comparison of the free energies illustrates Fig. \ref{free2.0}. This time the region $g\leq 2$ is inaccessible for this action \eqref{c1c32.0}.\\
The coordinates of the triple point are
\begin{align}
g_c\,&=\, +3.465045\pm0.000005\quad \nonumber\\
r_c\,&=\, -7.445855\pm0.000005\ .
\end{align}
\endminipage\hfill
\end{figure}

\subsubsection*{Phase diagram and triple point for ${t=-5}$}\label{}

The action \eqref{multiSm} with the parameter ${t=-5}$ is
\begin{align}
S(M)\,=\,\,\frac{1}{N}\slr{\frac12r\,\trl{M^2}+g\,\trl{M^4}}-5c_1c_3\ .\label{c1c35.0}
\end{align}
\vspace{-1cm}
\begin{figure}[h]
\minipage{0.49\textwidth}
    \centering
    \includegraphics[width=\linewidth]{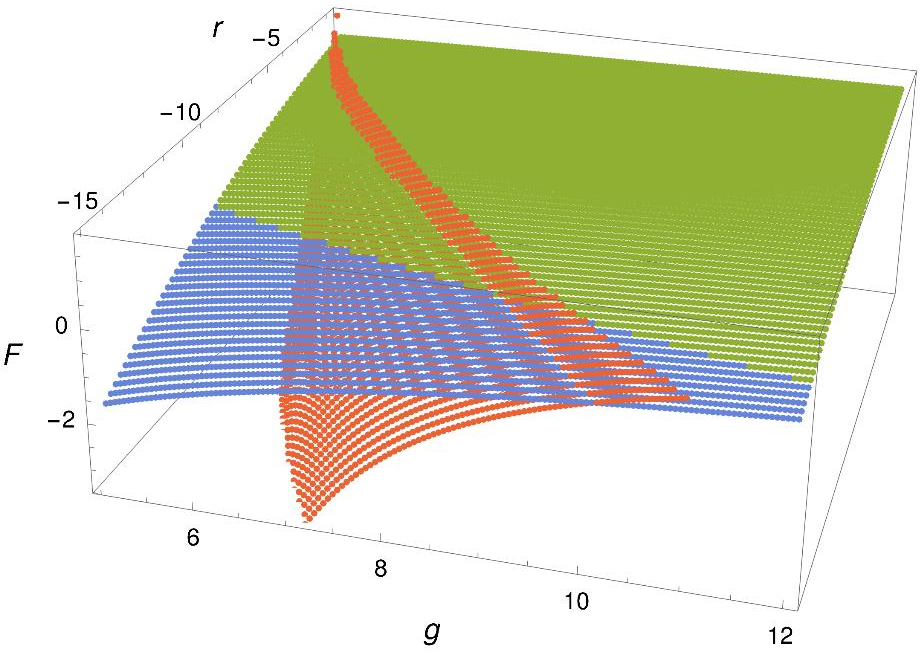}
    \caption{The free energies of the three phases of the model \eqref{c1c35.0} with the notation from Fig. \ref{free0}.}
    \label{free5.0}
\endminipage\hfill
\minipage{0.48\textwidth}

The correspondent comparison of the free energies illustrates Fig. \ref{free5.0}. This time the region $g\leq 5$ is inaccessible for this action \eqref{c1c35.0}.\\
The coordinates of the triple point are
\begin{align}
g_c\,&=\, +8.662625\pm0.000005\quad \nonumber\\
r_c\,&=\, -11.772935\pm0.000005\ .
\end{align}
\endminipage\hfill
\end{figure}

\section{Response functions}\label{sec4}

In section \ref{sec3} we first examined the action \eqref{multiSm} for a general $t$, than for four different values. From now on, we will be interested exclusively in the case ${t=-1}$, thus in the action \eqref{c1c3}.

In the previous two sections, we have described the solutions to saddle point equation of our system, and always having taken into account the most probable solution, we presented the phase diagram of the action \eqref{c1c3}. However, we have no insight to the inner behaviour of our model experimentally. So, we recall the response functions from statistical mechanics, which assign measurable physical quantities to the most probable microstates.

\subsection{Heat capacity $C$}
Let us introduce the temperature-dependent version of the action \eqref{c1c3}
\begin{align}
S(M, \beta)\,=\,\beta\,\lr{\frac{1}{N}\slr{\frac12r\,\trl{M^2}+g\,\trl{M^4}}-c_1c_3}\ ,\label{}
\end{align}
where $\beta$ is the inverse temperature.
The heat capacity $C$ (the analogy of the infinitesimal increase of energy caused by an infinitesimal increase of temperature for constant volume) is calculated from the analogy with statistical mechanics as
\begin{align}
C\,=\,-\eval{\frac{\partial^2}{\partial\beta^2}\,\F\lr{\beta}}_{\beta=1}\ .\label{cv}
\end{align}
By the redefinition of the parameters ${r\rightarrow\beta r}$ and ${g\rightarrow\beta g}$ for the symmetric regions without the effect of the multitrace term in \eqref{sec2.2}, we get a clear dependence in the formulae for the free energy from $\beta$, so the second derivative can be computed analytically, and at the end set ${\beta=1}$. Thus, we get
\begin{align}
\quad\text{symmetric 1-cut}\quad &C_{\text{S1}}=\frac{864\, g^2+r^4+r\sqrt{48\,g+r^2}\lr{24\,g -r^2 }}{3456\,g^2}\ ,\nonumber\\
\quad\text{symmetric 2-cut}\quad&C_{\text{S2}}\,=\,\frac{1}{4}\ . \label{cvs}     
\end{align}
In the asymmetric 1-cut case, we do not have an analytic relation for the free energy. So our idea to redefine the parameters is a cul-de-sac: since one does not have a \textcolor{black}{redefinable} parameter in front of the multitrace term and the scaling of the eigenvalues is not possible.
So, we are left with the only option to obtain the heat capacity, i.e. numerically
\begin{align}
C_{\text{A1}}\,=\,-\frac{\F\lr{1+\varepsilon}-2\F\lr{1}+\F\lr{1-\varepsilon}}{\varepsilon^2}\ .\label{cva}
\end{align}
This means that we calculate the free energy for a given point  $(g,\,r)$ for the three different values ${\beta\in\{1-\varepsilon,\,1,\,1+\varepsilon\}}$ \textcolor{black}{via the formula \eqref{Feff}} and we use them into the formula \eqref{cva}. \textcolor{black}{We have compared this numerical method with the formulae for the symmetric cuts \eqref{cvs}, and we got back the exact same results within the precision governed by $\varepsilon$, which was set ${\varepsilon=0.0001}$}. Furthermore, we have recovered the same behaviour of the heat capacity for ${g=1.1}$ and ${g=4}$ as acquired in \cite{ramda} by Monte Carlo simulations.\\

\begin{figure}[h]
    \centering
    \includegraphics[width=1.0\linewidth]{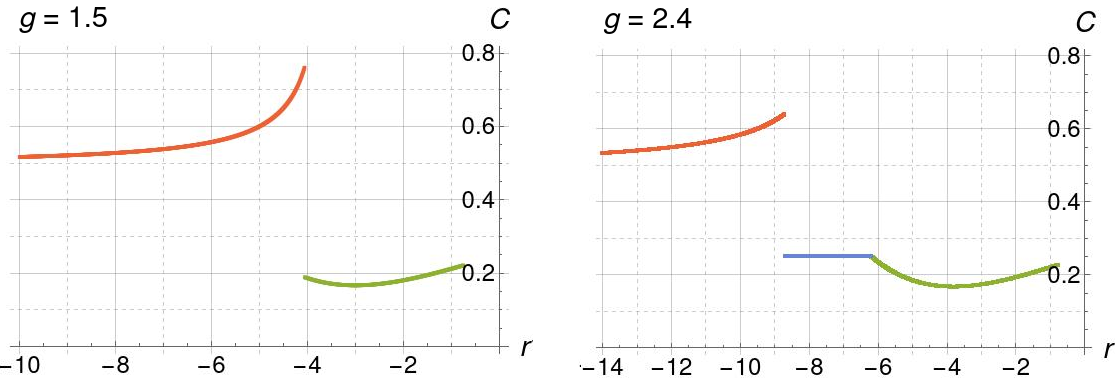}
    \caption{The heat capacity of the stable phases of the model \eqref{c1c3}.\\
    The notation is  the same as in the case of Fig. \ref{free0}.}
    \label{cv0}
\end{figure}

The function of the heat capacity $C$ was examined numerically for two values of $g$, the acquired graphs are in Fig. \ref{cv0}.

It is clear from the graphs that the phase transition of the heat capacity between the asymmetric 1-cut and the symmetric cuts does not show any sign of divergence in the vicinity of the transition point. A careful numerical investigation of the behavior of the heat capacity in the vicinity of the transition point also showed that the critical exponent is 0.

The graphs Fig. \ref{free0} and Fig. \ref{cv0} are useful for determining the type of the phase transition. In Fig. \ref{free0}, we see that the free energy of the system is a continuous function. However, we have proved numerically that in the vicinity of the phase transition line between the asymmetric 1-cut and the symmetric cuts, the first derivative of $\F$ with respect to $\beta$ is discontinuous, so the system suffers first-order phase transition. Similarly, in the right graph of Fig. \ref{cv0}, we observe that the heat capacity between the symmetric 1-cut and the symmetric 2-cut would have discontinuous derivative with respect to $\beta$ \textcolor{black}{(which we proved that happens indeed)}, so this is a third-order phase transition.

\subsection{Magnetization $m$}

The magnetization $m$ is calculated as
\begin{align}
m\,=\,\Bigg\langle\lim_{N\rightarrow\infty}\frac{1}{N}\sum_{i=1}^{N}\lambda_i\Bigg\rangle\,=\,\int_\C \mathrm{d}\lambda\,\lambda\,\rho(\lambda)\ ,\label{m}
\end{align}
see \cite{ydri}. Hereby, we are interested exclusively in the first moment of the distribution which satisfies the saddle point equation. For the symmetric phases, the odd moments vanish and thus so does the magnetization, and in the case of the asymmetric case we get the magnetization from the numerical calculations,
\begin{align}
m_{\text{A1}}\,=\, c_1\ .\label{ma}
\end{align}

There is a second independent approach, whence the magnetization can be obtained. Let us introduce the magnetic field-dependent version of the original action \eqref{c1c3}
\begin{align}
S(M, h)\,=\,\frac{1}{N}\slr{h\,\trl{M}+\frac12r\,\trl{M^2}+g\,\trl{M^4}}-c_1c_3\ ,\label{acth}
\end{align}
where $h$ is the magnetic field coupled to our matrix model, see \cite{ydri}.
So the magnetization $m$ can be calculated from the analogy with statistical mechanics as
\begin{align}
m\,=\,\eval{\frac{\partial}{\partial h}\,\F\lr{h}}_{h=0}\ .\label{mF}
\end{align}
The numerical version is
\begin{align}
m_{}\,=\,\frac{\F\lr{0+\varepsilon}-\F\lr{0}}{\varepsilon}\ ,\label{maa}
\end{align}
which means that we calculate the free energy for a given point  $(g,\,r)$ for two different values ${h\in\{0,\,0+\varepsilon\}}$ and we use them in the formula \eqref{maa}. 

Thus we have two independent approaches to calculate the magnetization. We have compared the numerical results of these two different paths, and we got back the exact same results within the precision governed by $\varepsilon$, which was set ${\varepsilon=0.0001}$.

For the sake of clarity, let us recall that the regions with the symmetric cuts were left untouched by the multitrace term, \textcolor{black}{so the magnetization of these values was automatically 0. For ${h\neq0}$ the stable solution was a slightly asymmetrized version of the originally symmetric solution, which means that the correspondent free energies were calculated by the general formulae for the 1-cut \eqref{F} and the 2-cut \eqref{F2} free energies.}

\begin{figure}[h]
    \centering
    \includegraphics[width=\linewidth]{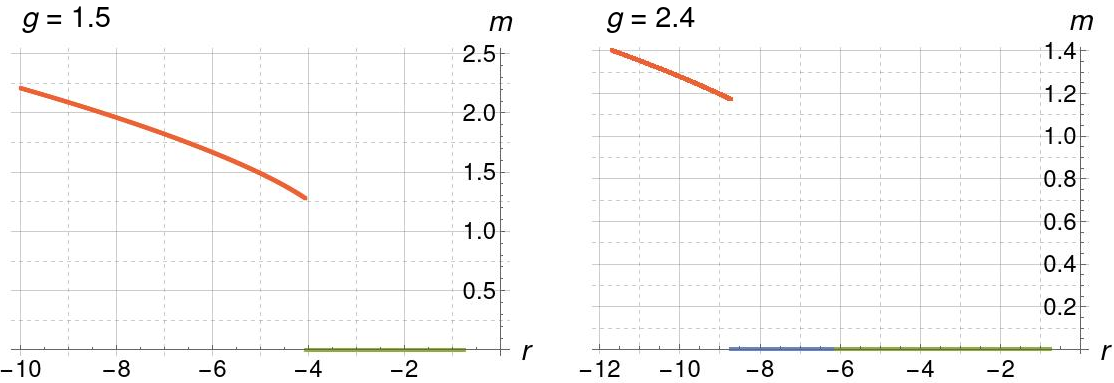}
    \caption{The magnetization of the stable phases of the model \eqref{c1c3}.\\
    The notation is  the same as in the case of Fig. \ref{free0}.}
    \label{m0}
\end{figure}

The function of the magnetization $m$ was examined for two values of $g$, the acquired graphs\footnote{Depending on whether the stable asymmetric 1-cut solution was found in the right or the left well of the potential, the first moment and thus the magnetization can be both negative or positive. For this reason some authors put the formula for magnetization \eqref{m} and \eqref{mF} into absolute value. But in this case problem would arise with the definition of the magnetic susceptibility \eqref{chiF} in the next section. In Fig. \ref{m0} we have illustrated only the positive values of the magnetization.} are in Fig. \ref{m0}.

\subsection{Magnetic susceptibility $\chi$}

The magnetic susceptibility $\chi$ (the analogy to the correlation function of the spins) is the correlation function of the eigenvalues, and by definition it is the variance  of the eigenvalues, 
\begin{align}
\chi\,=&\,\Bigg\langle\lr{\lim_{N\rightarrow\infty}\frac{1}{N}\sum_{i=1}^{N}\lambda_i}^2\Bigg\rangle-\Bigg\langle\lim_{N\rightarrow\infty}\frac{1}{N}\sum_{i=1}^{N}\lambda_i\Bigg\rangle^2\,=\,\langle\text{tr}^2\,(M)\rangle-\langle\text{tr}\,(M)\rangle^2\, .\label{chi}
\end{align}
The first term in \eqref{chi} cannot be obtained from the moments, so for further calculations we turn back to the action with the magnetic field \eqref{acth}. The magnetic susceptibility $\chi$ can be calculated from the analogy with statistical mechanics as
\begin{align}
\chi\,=\,-\eval{\frac{\partial^2}{\partial h^2}\,\F\lr{h}}_{h=0}\, .\label{}
\end{align}
The numerical version is \textcolor{black}{the second-order forward derivative}
\begin{align}
\chi\,=\,-\frac{\F\lr{0}-2\F\lr{0+\varepsilon}+\F\lr{0+2\varepsilon}}{\varepsilon^2}\, ,\label{chia}
\end{align}
which means that we calculate the free energy for a given point ${(g,\,r)}$ for the three different\footnote{ Practically, the numerical derivative \eqref{chia} is calculated for ${h=\varepsilon}$ and not for ${h=0}$. The reason why the derivative 
\begin{align*}
\chi\,=\,-\frac{\F\lr{0-\varepsilon}-2\F\lr{0}+\F\lr{0+\varepsilon}}{\varepsilon^2}\, ,\label{}
\end{align*}
is not working is the trait ${\F\lr{-h}=\F\lr{h}}$, which comes from the properties of the action \eqref{acth}. With the help of this identity the previous equation would become
\begin{align*}
\chi\,=\,-2\,\frac{\F\lr{\varepsilon}-\F\lr{0}}{\varepsilon^2}\,=\,-2\,\frac{m}{\varepsilon}\, ,\label{}
\end{align*}
which is in contradiction with the definition of the magnetization \eqref{maa} due to its clear dependence from $\varepsilon$. 
}
${h\in\{0,\,0+\varepsilon,\,0+2\varepsilon\}}$ and we insert them into the formula \eqref{chia}. 

If we take into account the formula for magnetization \eqref{maa}, we can rewrite the definition of the magnetic susceptibility 
\begin{align}
\chi\,=\,-\eval{\frac{\partial}{\partial h}\,m\lr{h}}_{h=0}\, ,\label{chiF}
\end{align}
whose numerical version is
\begin{align}
\chi\,=\,-\frac{m\lr{0+\varepsilon}-m\lr{0}}{\varepsilon}\, .\label{chia1}
\end{align}

The magnetic susceptibility as the correlation function of the eigenvalues should consider the model we are working in as a whole including the multitrace terms. In other words, the magnetic susceptibility of the symmetric cuts should be calculated when one takes into account the whole action \eqref{acth} for ${h\neq0}$. Thus, the multitrace term in the action \eqref{acth} modifies the magnetic susceptibility of the symmetric phases compared to their magnetic susceptibility in the pure potential case.

The equations \eqref{chia} and \eqref{chia1} represent two mathematical expressions of the exact same method. We have compared the numerical results of these two paths, and we got back the same results for the symmetric 1-cut and the asymmetric 1-cut regimes within the precision governed by $\varepsilon$, which was set ${\varepsilon=0.001}$. 

\begin{figure}[h]
    \centering
   \includegraphics[width=\linewidth]{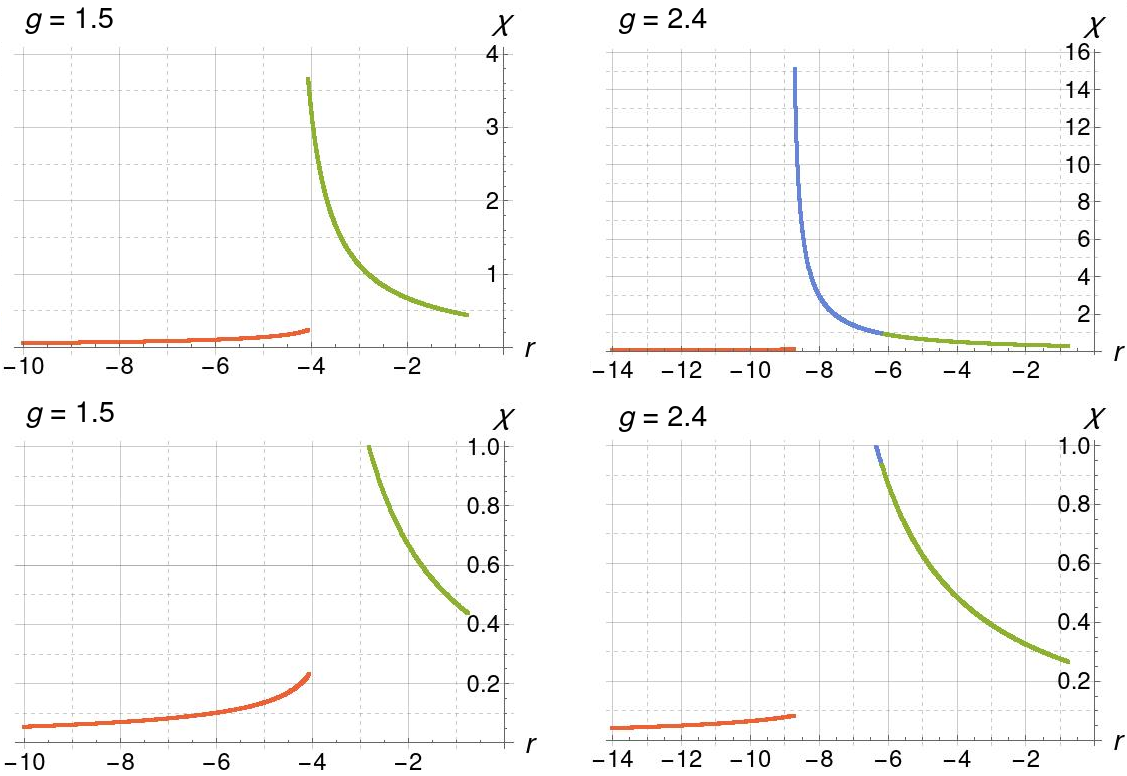}
   \caption{The magnetic susceptibility of the stable phases of the model \eqref{c1c3}.\\
    The notation is  the same as in the case of Fig. \ref{free0}.\textcolor{black}{ The upper two graphs show the whole functions of the magnetic susceptibility, while the lower two graphs show the zoomed functions of the magnetic susceptibility exclusively for values less than $1$.} }
   \label{chi0}
\end{figure}

In the case of the symmetric 2-cut solutions for ${h=0}$ we calculated the correspondent asymmetric 2-cut solution for ${h\neq0}$ in the way that we sliced up the interval ${[0,1]\ni k}$ from subsection \ref{desc} for 500 equidistant cuts and we searched for the asymmetric 2-cut solution with the lowest free energy. We have compared the numerical results of these two paths, and we got back the same results for the symmetric 2-cut within the precision of our parameters:
\begin{itemize}
  \item $\varepsilon$, which was set ${\varepsilon=0.01}$,
  \item the number of the examined values in the interval ${[0,1]}$, which was set as 500,
  \item and the fact that in real \eqref{chia} calculates the magnetic susceptibility for ${h=\varepsilon}$, whereas \eqref{chia1} calculates the magnetic susceptibility for ${h=\varepsilon/2}$.
\end{itemize}
In one hand, the relation \eqref{chia} is more sensible for the value of $\varepsilon$, while in the other hand, the relation \eqref{chia1} is more sensible for the number of the examined values in the interval ${[0,1]}$. In practice these two parameters should be varied in synch.
 The last mentioned bullet point became decisive near the transition point, where the magnetic susceptibility started to grow rapidly; the difference of the numerical values of these two methods was approximately around 1. The acquired graphs of the function of the magnetic susceptibility $\chi$  are in Fig. \ref{chi0}, the magnetic susceptibility for the symmetric 2-cut was calculated via \eqref{chia}.

Despite the growth, it is clear from the graphs in Fig. \ref{chi0} that the phase transition of the magnetic susceptibility between the asymmetric 1-cut and the symmetric cuts does not show any sign of divergence in the vicinity of the transition point. A careful numerical investigation of the behavior of the graphs of the magnetic susceptibility in the vicinity of the transition point also showed that the critical exponent vanishes.

It is visible from the Fig. \ref{free0} that directly before the transition point from the symmetric 2-cut phase to the asymmetric 1-cut phase, the asymmetric 1-cuts solutions come to existence, however, their free energies are greater than the symmetric 2-cuts'. In this interval where the asymmetric 1-cut solutions already exist, the stable asymmetric 2-cut solutions with ${h\neq0}$ for ${0<k\ll1}$ have the tendency to approach the asymmetric 1-cut, and the stable phase is realized for the value of $k$ closest to 0. Since this type of solution is not the correspondent one to our symmetric 2-cut solution with ${h=0}$, we did not use this stable solution. For values ${0\ll k<1}$ we still found a stable solution which had, however, greater free energy than the asymmetric 2-cut with $k$ close to 0. So, we refer to those solution with ${0\ll k<1}$ as metastable solutions, and their free energies were plugged in \eqref{chia}.

We have checked our numerical results for the asymmetric 1-cut regime also by performing numerical Monte Carlo simulations as well, and this numerical Monte Carlo analysis recovered the exact same numerical values as our study suggested. The graphs of the magnetic susceptibility acquired by Monte Carlo simulations for ${g=1.1}$ and ${g=4}$ were presented in \cite{ramda}. Qualitatively, we found the same behaviour of the magnetic susceptibility but quantitatively we collected at least one order higher results. We could not localize the exact step whence the discrepancy arises.

%\begin{comment}
\subsection{String susceptibility $\gamma$}
In the string theory, the string susceptibility describes the number of possible string configurations as the string lengthens. In matrix models related to quantum gravity.
%, the string susceptibility provides information about \textcolor{blue}{how} the discretized surfaces of the Feynman diagrams change their behaviour, see \cite{sumit}
The string susceptibility $\gamma$ is defined as
\begin{align}
\gamma\,=\,\frac{\partial^2}{\partial g^2}\,\F(g)\, .\label{}
\end{align}
%\end{comment}

\subsubsection*{String susceptibility for the multitrace term $c_2^2$}\label{}
In the subsection \ref{phasediag} we have already mentioned that for the action 
\begin{align}
S(M)\,=\,\,\frac{1}{N}\slr{\frac12r\,\text{tr}(M^2)+g\,\text{tr}(M^4)}+ g'\, c_2^2 \label{c22} 
\end{align}
for the values ${g'\in\{-1,\,1\}}$ we recovered the same phase diagram as the analytical approach suggests in \cite{jt15b, rus}. The string susceptibility of this model \eqref{c22} was examined in \cite{sumit, cicuta}. The article \cite{cicuta} points out that the way the string susceptibility was analyzed in \cite{sumit} was executed improperly due to some terms forgotten in the differentiation. The article \cite{cicuta} starts its calculation with their normalized free energy, however, it leaves out the multitrace contribution to the free energy and essentially takes $\mathcal{F}_{\text{eff}}$ instead of $\mathcal{F}$. The normalized free energy -- in our notation -- from \eqref{s1} is
\begin{align}
\mathcal{F}\,=\,-\frac12\ln{\lr{\frac{\delta}{4} r}}-\frac14 g \delta^2-\frac{3}{128}g^2\delta^4-g'\, c_2^2\,\text{, where}\,\,\, r>0\, . \label{} 
\end{align}
Using the relations of the derivative of the second moment with respect to $g$ derived in \cite{cicuta}, we got that the string susceptibility for this $g'c_2^2$ multitrace model
\begin{align}
\gamma\,=\,-\frac{\delta^4}{64}\frac{36+g'\delta^2}{4+g'\delta^2}\,  \label{ssusc2} 
\end{align}
as it was stated in \cite{sumit}. Our numerical analysis confirmed the correctness of the critical exponents of the string susceptibility at the phase transitions derived from the original free energy in \cite{cicuta}. However, the critical exponent of the string susceptibility \eqref{ssusc2} for the regime ${g'>-\frac{9}{256}r^2}$ should be $0$ in \cite{sumit}.

\subsubsection*{String susceptibility for the multitrace term $-c_1c_3$}\label{}

In the symmetric regions without the effect of the multitrace term in \eqref{sec2.2}, we got a clear dependence in the formulae for the free energy from $g$, so the second derivative can be executed analytically. Thus, we get
\begin{align}
\quad\text{symmetric 1-cut}\quad &\gamma_{\text{S1}}=\frac{ r\left(\sqrt{48 g+r^2}-r\right)\lr{24g+r^2}-24g\lr{12g+r^2}}{1152 g^4}\ ,\nonumber\\
\quad\text{symmetric 2-cut}\quad&\gamma_{\text{S2}}\,=\,\frac{2 g+r^2}{8 g^3}\ . \label{strings}     
\end{align}
In the asymmetric 1-cut case, we do not have an analytic relation for the free energy, so we are left with the numerical approach to obtain string susceptibility
\begin{align}
\gamma_{\text{A1}}\,=\,\frac{\F\lr{g+\varepsilon}-2\F\lr{g}+\F\lr{g-\varepsilon}}{\varepsilon^2}\ ,\label{stringa}
\end{align}
which means that we calculate the free energy for a given point  $(g,\,r)$ for the three different values ${\{g-\varepsilon,\,g,\,g+\varepsilon\}}$ and we insert them into the formula \eqref{stringa}. We have compared this numerical method with the formulae for the symmetric cuts \eqref{strings}, and we got back the same results within the precision governed by $\varepsilon$, which was set ${\varepsilon=0.001}$. \\

\begin{figure}[h]
    \centering
    \includegraphics[width=1.0\linewidth]{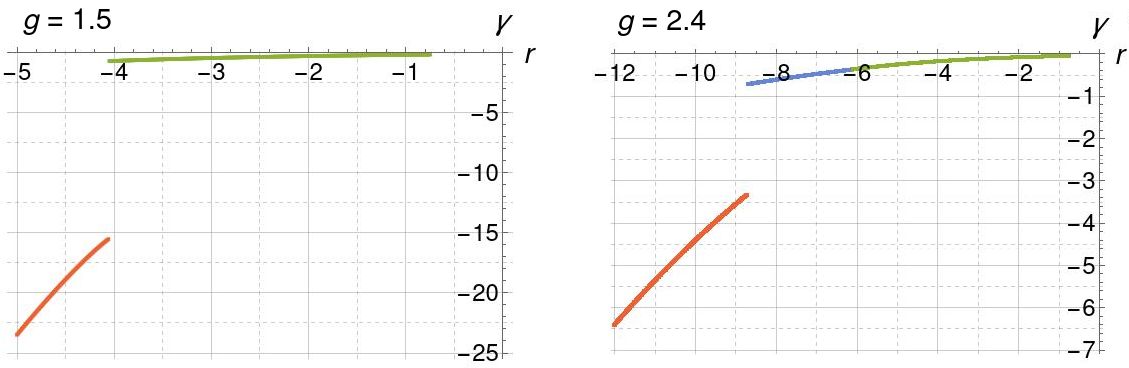}
    \caption{The string susceptibility of the stable phases of the model \eqref{c1c3}.\\
    The notation is  the same as in the case of Fig. \ref{free0}.}
    \label{string0}
\end{figure}

The function of the string susceptibility $\gamma$ was examined numerically for two values of $g$, the acquired graphs are in Fig. \ref{string0}. 

\section{Conclusion}\label{sec5}
In this work, we have enriched the quartic potential with a multitrace term, resulting in the action \eqref{c1c3}. This is the second simplest multitrace matrix model which stabilizes the asymmetric 1-cut solution, after the multitrace term of type\footnote{Models with this kind of multitrace interaction have been analyzed in \cite{msjt}.} $-c_1^2$. For the intact symmetric regions the analytic solutions were used, the numerical treatment was applied in the case of the asymmetric 1-cut solution. We have obtained the phase diagram, and we have found the triple point of this model.

Besides that, we have probed the response functions of this model. The heat capacity was calculated with the help of the introduced inverse temperature, the magnetisation and the magnetic susceptibility were calculated with the help of the introduced magnetic field. The string susceptibility was first recalculated in a model defined by the multitrace term $-g'\,c_2^2$ before investigating it for our model.

It is an intriguing observation that neither of the response functions showed divergent behavior in the vicinity of the transition point between the symmetric 1-cut and the asymmetric 1-cut regimes or the symmetric 2-cut and the asymmetric 1-cut regimes.

The examined mere action \eqref{c1c3} does not describe any physical system so far. The only meaningful interpretation would be to consider the multitrace term $-c_1 c_3$ as a kinetic term of a theoretical matrix model. However, by proper understanding of this model, we gain precious insight to the analysis of the multitrace terms. For instance, the interaction term of the type ${-c_1 c_3-c_2^2}$ describes a type of random fuzzy geometries \cite{john,khlakhali,randomNC}, and a much more complicated multitrace term describes the second order kinetic term effective action of the Grosse-Wulkenhaar model \cite{svk2023}
\begin{align}
S_\text{kin}^\text{eff}&=N^2 \slr{\lr{c_2-c_1^2}-\frac{1}{120}\left(565 c_1^4-1130 c_1^2 c_2+388 c_1 c_3+274 c_2^2-97 c_4\right)}\ .\label{GWkin}
\end{align}

One of the most interesting aspects of the simple multitrace models is the possible stability of the asymmetric 2-cut solutions. Here, we have checked that this particular model does not have such regime and as soon as the model decides to go asymmetric it goes all the way to 1-cut solution. This was a very tedious and demanding task, however it brings a lot of insight into the problem. We hope to be able to investigate existence of the asymmetric 2-cut solutions in other models soon, as there are hints of existence of such solutions in models describing random geometries and also some more complicated fuzzy field theories \cite{asym2cut}.

Finally let us mention that the bootstrap program, see e.g. \cite{randomNC,Khalkhali:2023onm,Hessam:2021byc}, is emerging as a very powerful tool to analyze various matrix models. It would be interesting to see, which approach is better to answer what type of questions.

\paragraph{Acknowledgments.}
We would like to express our special thanks to Samuel Kováčik and Dragan Prekrat for checking the numerical results of the magnetic susceptibility and reading through our paper, to Matej Hrmo for checking the numerical results of the heat capacity from the article and to Giovanni Cicuta for bringing our attention to the article \cite{cicuta}. \\
This work has been supported by VEGA 1/0025/23 grant \emph{Matrix models and quantum gravity} and the Collegium Talentum Programme of Hungary.

%\bibliography{sn-bibliography}

\end{document}